\documentclass[referee]{raa}            % referee version: for submission

\usepackage{graphicx,times}             %for PS/EPS graphics inclusion, new
\usepackage{threeparttablex}
\usepackage{natbib}
\usepackage{textcomp}
\usepackage{mathtext}
\usepackage{amssymb,amsmath}
\usepackage{longtable,booktabs}
\newcommand{\fermi}{\textit{Fermi}}
\newcommand{\gr}{$\gamma$-ray}
\newcommand{\first}{PSR~J1017$+$30}
\newcommand{\second}{PSR~J1715$+$46}

\begin{document}

   \title{Searching for $\gamma$-ray emission from LOTAAS pulsars} 
%\footnotetext{$*$ Supported by the National Natural Science Foundation of China.}
%   \subtitle{I. Place Your Subtitle Here}

   \volnopage{Vol.0 (2015) No.0, 000--000}      %%preserved for Editor. DOn't remove!
   \setcounter{page}{1}          %%starting page, preserved for Editor. DOn't remove!

   \author{Qi-Wei Lu 
      \inst{1,2}
   \and Zhong-Xiang Wang
      \inst{3,1}
   \and Yi Xing
      \inst{1}
   }

   \institute{Shanghai Astronomical Observatory, Chinese Academy of Sciences,
             Shanghai 200030, China; {\it wangzx@shao.ac.cn}\\
        \and
             Graduate University of the Chinese Academy of Sciences, Beijing 100049, China\\
	\and
	Department of Astronomy, Yunnan University, Kunming 650091, China\\
   }

%%   \date{Received~~2009 month day; accepted~~2009~~month day}

\abstract{The LOw-Frequency ARray (LOFAR) has recently conducted a survey 
(LOFAR Tied-Array All-sky Survey; LOTAAS) for
pulsars in the Northern hemisphere that resulted in discoveries of 73 new 
pulsars. 
For the purpose of studying the properties of these pulsars, we search for 
their \gr\
counterparts using the all-sky survey data obtained with the Large Area 
Telescope (LAT) onboard {\it the Fermi Gamma-Ray Space Telescope (Fermi)}.
We analyze the LAT data for 70 LOTAAS pulsars (excluding two millisecond pulsars
and one with the longest known spin period of 23.5~s). We find one candidate
counterpart to \first, which should be searched for the \gr\
pulsation signal once its timing solution is available. 
For other LOTAAS pulsars, we derive their 0.3--500~GeV
flux upper limits. In order to compare the LOTAAS pulsars with 
the known \gr\ pulsars, we also derive the 0.3--500\,GeV \gr\ fluxes 
for 112 
of the latter contained in the \textit{Fermi} LAT fourth source catalog.
Based on the properties of the \gr\
pulsars, we derive upper limits on the spin-down luminosities of the LOTAAS
pulsars. The upper limits are not very constraining but help suggest that
most of the LOTAAS pulsars probably have $< 10^{33}$\,erg\,s$^{-1}$ spin-down
luminosities and are not expected to be detectable with \fermi\ LAT.
}

   \authorrunning{Q. Lu, Z. Wang, \& Y. Xing}            %author_head in even pages
   \titlerunning{Searching for $\gamma$-ray emission from LOTAAS pulsars}  % title_head in odd pages

   \maketitle
%% The author head (on even pages) and the title head (on odd pages) will be
%% automatically extracted from \author{} and \title{}. Whenever the title is too long,
%% you will be asked to supply a shorter one by inserting either \authorrunning{} or
%% \titlerunning{} before \maketitle. Anyway, you can specify your own heads.
%%
%%
%% Note: In the following text body of your manuscript, please note several differences from
%%       other major journals:
%% (1) \subsection{Please Capitalize the First Letter of Each Notional Word in Subsection Title}
%% (2) Please Capitalize the First Letter of Each Notional Word in all tables' captions

%
%________________________________________________ sections below
%
\section{Introduction}           %% first-level sections will be auto-capitalized
\label{sect:intro}

Since the discovery of the first radio pulsar signal in 1967 \citep{hew+68},
nearly 3000 pulsars have been found \citep{man+05}, resulting from different
surveys mainly at radio bands. Among them, approximately 2000 are ``young" radio
pulsars (in this paper, ``young" pulsars are used to distinguish them
from most of the others, the old ``recycled" millisecond pulsars; MSPs).
These pulsars are believed to be born in supernova explosions, while in total 
there are possibly $\sim$100,000 of them in our Galaxy 
(e.g., \citealt{swi+14}).
The new generation radio telescopes such as the Five-hundred-meter Aperture 
Spherical Telescope (FAST) and the near-future Square Kilometer Array (SKA)
will potentially be able to find and study most of these pulsars
(e.g., \citealt{qia+19,kea+15}) and thus lead to a more complete 
understanding
of the physical properties of the radio pulsar population.

Recently, a survey for radio pulsars in the Northern hemisphere at the very
low frequency range of 119--151~MHz was carried out with the LOw-Frequency 
ARray (LOFAR; \citealt{van+13}). This LOFAR Tied-Array All-sky Survey 
(LOTAAS) has resulted in the discovery of 73 new radio pulsars \citep{san+19}.
These LOTAAS pulsars probably represent a sample of pulsars that are
bright at low frequencies and
are found to have longer spin periods than the known young pulsar 
population.  It is thus interesting to study their overall properties, checking
if there are any other differences between them and the known young pulsar 
population or other pulsar groups based on survey methods.

From the beginning of the \gr\ astronomy, it has been learnt
and pointed out that pulsars are potential high-energy objects with 
$\gamma$-rays emitted from the magnetosphere due to their high
surface magnetic fields
(e.g., \citealt{ry95}). Observations conducted with 
{\it The Fermi Gamma-Ray Space Telescope (Fermi)}, which was launched in 2008, 
have confirmed that pulsars are the dominant 
\gr\ sources in our Galaxy \citep{2fpsr13}. The Large Area Telescope (LAT)
onboard \fermi\ has been scanning the whole sky at the energy band
of 0.1--500 GeV and thus far 253 pulsars, according to the latest results
provided by the \fermi\ LAT Multiwavelength Coordinating Group (MCG)$\footnote{https://confluence.slac.stanford.edu/display/GLAMCOG/Public+List+of+LAT-Detected+Gamma-Ray+Pulsars}$, have been detected with pulsed \gr\ emission.
Among them, 135 are young pulsars. The \gr\ properties of these pulsars 
allow a deep probe into the emission mechanism of pulsars 
(e.g., \citealt{pie+12}).

Taking advantage of the all-sky \gr\ data collected by LAT, we carried out
the search for \gr\ counterparts to the LOTAAS pulsars. The results of any
detection or upper limits provide constraints on properties of this very 
low-frequency pulsar sample. In this paper, we report the results from
our search.  Below in Section 2,  we describe the analysis of the \fermi\ LAT
data and provide the results. In Section 3, we discuss the implication
of the results.
    
\section{LAT Data Analysis and Results}
\label{sec:oda}

\subsection{LAT data and source models}
LAT is a $\gamma$-ray imaging instrument that continuously scans the whole 
sky in the GeV band \citep{atw+09}. In the analysis, we selected 0.1--500 GeV 
LAT events inside a $\mathrm{20^{o}\times20^{o}}$ region centered at the 
position of each of our targets.  The time period of the LAT data was
from 2008-08-04 15:43:36 (UTC) to 2019-08-14 02:23:15 (UTC). 
The updated \textit{Fermi} Pass 8 database was used.
Following the recommendations of the LAT 
team\footnote{\footnotesize http://fermi.gsfc.nasa.gov/ssc/data/analysis/scitools/}, 
we excluded the events with zenith angles larger than 90 degrees (to prevent 
the Earth's limb contamination) and with quality flags of `bad'.

For each of the targets, we constructed a source model. 
The sources, which are listed in the 
\textit{Fermi} LAT fourth source catalog 
(4FGL; \citealt{4fgl20}) and within a 20 degree radius circular 
region from a target, were included in the source model. 
The spectral forms of the sources are provided in 4FGL. 
In our analysis, we set the spectral parameters of the sources 
within 5 degrees of a target as free parameters, and fixed
the other parameters at their catalog values. 
The background Galactic and extragalactic diffuse 
spectral models gll\_iem\_v07.fits and the file iso\_P8R3\_SOURCE\_V2\_v1.txt, 
respectively, were also included in the source model, and
the normalizations of the two models were set as free parameters.

\subsection{LOTAAS pulsars}

Among the 73 radio pulsars discovered by LOTAAS, two are MSPs and 
PSR J0250+58 has the longest spin period ($P=23.5$\,s) among the known pulsars
\citep{san+19}. We excluded these three pulsars from our target list; 
the former are presumably exceptions in the LOTAAS pulsar sample 
and the latter was studied in detail by \citet{tan+18}. 
Assuming power-law emission for each of the LOTAAS pulsars in the source models 
at their
positions given in \citet{san+19}, we performed standard binned likelihood 
analysis of the LAT data using {\tt Fermitools}. 
To avoid the relatively large uncertainties of the instrument response 
function of LAT and the strong background emission (or possible contamination
from nearby sources) in the Galactic plane in the low energy range 
of $<0.3$~GeV, we included events in the energy range of 0.3--500~GeV for 
the likelihood analysis. We re-fit the source models using \textit{gtlike}
to the LAT data. With the fitted source models,
the Test Statistic (TS) map of a $\mathrm{3^{o}\times3^{o}}$ region centered 
at each of the pulsar targets was calculated (using \textit{gttsmap}). 
All the catalog sources were included in 
the source models. A TS value at a given position 
is a measurement of the fit improvement for including 
a source at the position, and is approximately the square of the detection 
significance of the source \citep{4fgl20}.
From the TS maps, we found only two possibly detected sources near the positions
of \first\ and \second. For the other 68 targets without any significantly
detected sources at their positions, we derived 95\% flux upper limits, which
are given in Table~\ref{tab1}.
\begin{longtable}{lccc}
%%\centering
%%\small
%%	\begin{minipage}[]{100mm}
\caption{0.3--500 GeV flux upper limits of 68 LOTAAS pulsars.}
\label{tab1}\\
 %%      \end{minipage}
%%%	\fontsize{6}{6}\selectfont
	%%\begin{tabular}{lccc}
		\hline
		Source & R.A. & Decl. & Flux/10$^{-13}$  \\	
		name & (h:m) & ($^{\circ}$:$'$) & (erg\,cm$^{-2}$\,s$^{-1}$) \\
		\hline 
\endfirsthead
\hline
Source & R.A. & Decl. & Flux/10$^{-13}$  \\
                name & (h:m) & ($^{\circ}$:$'$) & (erg\,cm$^{-2}$\,s$^{-1}$) \\
\hline
\endhead
\hline
\endfoot
		J0039$+$35 & 00:39.1&+35:45&6.3 \\
		J0059$+$69&00:59.5&+69:55&14 \\
		J0100$+$80&01:00.3&+80:22&4.4 \\
		J0107$+$13&01:07.6&+13:25&3.1 \\
		J0115$+$63&01:15.6&+63:24&6.1 \\
		J0121+14  & 01:22.0 & 14:16 & 8.4\\
		J0139$+$33&01:40.0&+33:37&2.0 \\
		J0210$+$58&02:11.0&+58:44&11 \\
		%%J0250+58&02:50.3&+58:54&9.5  \\
		J0302$+$22&03:02.5&+22:50&1.5 \\
		J0305$+$11&03:05.1&+11:23&4.9 \\
		J0317$+$13&03:17.9&+13:29&5.3 \\
		J0349$+$23&03:49.9&+23:41&15 \\
		J0421$+$32&04:21.4&+32:54&17 \\
		J0454$+$45&04:54.9&+45:28&15 \\
		J0518$+$51&05:18.3&+51:25&4.1 \\
		J0742$+$43&07:42.6&+43:33&1.9 \\
		J0811$+$37&08:11.2&+37:28&1.8 \\
		J0813+22 & 08:13.9 & 22:01 & 2.2 \\
		J0857$+$33&08:57.8&+33:48&4.5 \\
		J0928$+$30&09:29.0&+30:38&3.9 \\
		J0935$+$33&09:35.1&+33:11&0.88 \\
		%%J1017$+$30&10:17.6&+30:10&12 \\
		J1226$+$00&12:26.2&+00:03&7.1 \\
		J1235$-$02&12:35.9&-02:05&5.8 \\
		J1303$+$38&13:03.3&+38:13&4.1 \\
		J1334$+$10&13:34.5&+10:05&3.3 \\
		J1344$+$66&13:43.9&+66:33&14 \\
		J1404$+$11&14:04.6&+11:57&2.6 \\
		J1426$+$52&14:27.0&+52:10&1.7 \\
		J1529$+$40&15:29.2&+40:49&9.4 \\
		J1623$+$58&16:23.8&+58:49&1.3 \\
		J1635$+$23&16:35.1&+23:31&1.8 \\
		J1638$+$40&16:38.8&+40:05&8.5 \\
		J1643$+$13&16:43.8&+13:25&3.8 \\
		J1655$+$62&16:55.9&+62:02&2.0 \\
		J1657$+$33&16:57.7&+33:03&6.1 \\
		J1707$+$35&17:07.0&+35:56&6.5 \\
		J1713$+$78&17:13.5&+78:09&2.2 \\
		%%J1715$+$46&17:15.8&+46:03&26 \\
		J1722$+$35&17:22.1&+35:18&3.9 \\
		J1735$+$63&17:35.1&+63:19&2.3 \\
		J1740$+$27&17:40.5&+27:13&8.5 \\
		J1741$+$38&17:41.2&+38:54&7.4 \\
		J1745$+$12&17:45.7&+12:51&3.3 \\
		J1745$+$42&17:45.8&+42:53&8.9 \\
		J1749$+$59&17:49.6&+59:51&1.5 \\
		J1809$+$17&18:09.1&+17:04&15 \\
		J1810$+$07&18:10.7&+07:03&2.9 \\
		J1814$+$22&18:14.6&+22:23&12 \\
		J1848$+$15&18:48.9&+15:17&4.9 \\
		J1849$+$25&18:49.8&+25:58&14 \\
		J1910$+$56&19:10.7&+56:55&0.82 \\
		J1916$+$32&19:16.1&+32:24&6.8 \\
		J1933$+$53&19:33.0&+53:32&2.7 \\
		J1953$+$30&19:53.8&+30:13&3.0 \\
		J1957$-$00&19:57.6&$-$00:01&2.3 \\
		J1958$+$56&19:58.0&+56:49&8.3 \\
		J2006$+$22&20:06.6&+22:04&3.2 \\
		J2022$+$21&20:22.4&+21:11&2.3 \\
		J2036$+$66&20:36.8&+66:44&8.3 \\
		J2051$+$12&20:51.4&+12:48&2.8 \\
		J2053$+$17&20:53.8&+17:18&16 \\
		J2057$+$21&20:57.8&+21:26&14 \\
		J2122$+$24&21:22.7&+24:24&11 \\
		J2123$+$36&21:23.8&+36:24&3.4 \\
		J2209$+$22&22:09.9&+21:17&1.6 \\
		J2306$+$31&23:06.2&+31:23&6.1 \\
		J2329$+$47&23:29.6&+47:42&2.7 \\
		J2336$-$01&23:36.6&$-$01:51&12 \\
		J2350$+$31&23:50.7&+31:39&4.9 \\
	\hline	
	%%\end{tabular}
\end{longtable}

\subsubsection{PSR J1017+30}

\begin{figure}
   \centering
\includegraphics[width=0.49\textwidth]{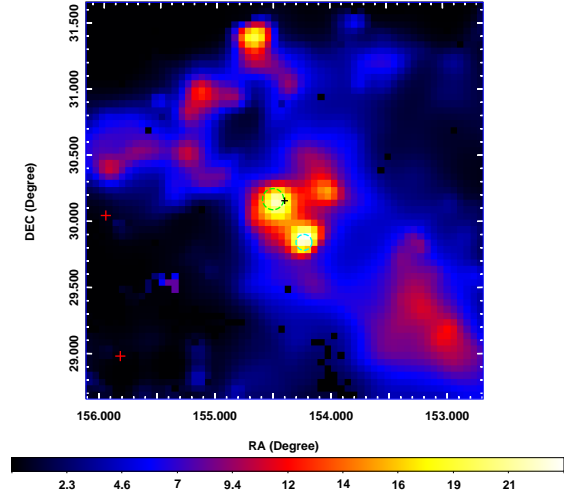}
\caption{0.3--500 GeV TS map of the $\mathrm{3^{o}\times3^{o}}$ region centered 
at \first. The image scale of the map is 0\fdg05 pixel$^{-1}$. All 4FGL catalog
sources (red pluses) were considered and removed.  
The black plus indicates the radio position with the length being 
the uncertainty of 3$'$. The two circles are the 2$\sigma$ error circles of 
the best-fit positions obtained for the two sources from \fermi\ analysis. 
The northeast source has a position consistent
with that of \first.
}
\label{fig:j1017}
\end{figure}

There are two \gr\ sources, clearly separated with each other, near 
the position of \first\ with TS 
values of $\sim$22 (Figure~\ref{fig:j1017}). We ran \textit{gtfindsrc} in 
{\tt Fermitools} to determine their positions and obtained 
R.A.=154\fdg50, Decl.=30\fdg18 (equinox J2000.0) for the northeast (NE) one 
and R.A.=154\fdg23, Decl.=29\fdg85 (equinox J2000.0) for the southwest (SW) 
one. The 1$\sigma$ nominal uncertainties are 0\fdg05 and 0\fdg04, respectively.
\first\ is 0\fdg09 away from the position of the NE source, but given 
the positional
uncertainty of 3$'$ in the LOTAAS survey \citep{san+19}, the two sources match
in position (see Figure~\ref{fig:j1017}), i.e., the NE source could be the  
\gr\ counterpart to \first.

Including the NE and SW sources in the source model,
we re-performed the likelihood analysis to the 0.3--500 GeV data, in which
a power law was assumed for the two sources.
Note that pulsars' \gr\ emission
can generally be described with an exponentially cutoff power law
\citep{2fpsr13}, but since the detection significance of the NE source was
low, we chose to use a simple power law instead (we tested the former
model, but no higher significant results were obtained).
We obtained $\Gamma= 2.4\pm$0.3 and 
$F_{0.3-500}= 4\pm2\times 10^{-10}$ photons~s$^{-1}$\,cm$^{-2}$ for 
the NE source with a TS value of 15 ($\Gamma= 1.9\pm$0.3,
$F_{0.3-500}= 2\pm1\times 10^{-10}$ photons~s$^{-1}$\,cm$^{-2}$ for 
the SW source with a TS value of 16). We also tested to use the data from
0.1~GeV, $\Gamma$ was nearly the same but the flux was increased to
$2\pm1\times 10^{-9}$ photons~s$^{-1}$\,cm$^{-2}$ and TS$\simeq 19$
Therefore the NE source was detected at $\simeq 4\sigma$.

We extracted the \gr\ spectra of the NE source by performing maximum 
likelihood analysis of the LAT data in 10 evenly divided energy bands in 
logarithm from 0.1--500 GeV. 
In the extraction, the spectral normalizations of the sources within 5 
degree from it were set as free parameters, while all the other parameters 
of the sources were fixed at the values obtained from the above maximum 
likelihood analysis. 
For the results, we kept only spectral data points when TS is greater 
than 5 ($>$2$\sigma$ significance) and derived 95\% flux upper limits 
otherwise. 
The flux and TS values of the spectral data points are provided in 
Table~\ref{tab:spectra}. 
\begin{table}
\begin{center}
\caption{\fermi\ LAT flux measurements for the sources at the positions of \first\ and \second.}
\label{tab:spectra}
\begin{tabular}{lccccc}
\hline
\multicolumn{2}{c}{ } &
\multicolumn{2}{c}{\first} &
\multicolumn{2}{c}{\second} \\ \hline
$E$ & Band & $F/10^{-12}$ & TS & $F/10^{-12}$ & TS \\
(GeV) & (GeV) & (erg cm$^{-2}$ s$^{-1}$) &  & (erg cm$^{-2}$ s$^{-1}$) &  \\ \hline
0.15 & 0.1--0.2 & 0.9$\pm$0.5 & 9 & 1.6 & 2 \\
0.36 & 0.2--0.5 & 0.8 & 2 & 0.7 & 4 \\
0.84 & 0.5--1.3 & 0.3 & 2 & 0.2 & 0 \\
1.97 & 1.3--3.0 & 0.2 & 2 & 0.2 & 2 \\
4.62 & 3.0--7.1 & 0.14$\pm$0.08 & 5 & 0.1 & 0 \\
10.83 & 7.1--16.6 & 0.2 & 1 & 0.2 & 0 \\
25.37 & 16.6--38.8 & 0.4 & 3 & 0.5$\pm$0.2 & 17 \\
59.46 & 38.8--91.0 & 0.3 & 0 & 0.3 & 0 \\
139.36 & 91.0--213.3 & 0.8 & 0 & 2.0 & 2 \\
326.60 & 213.3--500.0 & 1.9 & 0 & 1.6 & 0 \\
\hline
\end{tabular}
\\
\footnotesize{Note: $F$ is the energy flux ($E^{2} dN/dE$).  Fluxes without uncertainties are 
the 95$\%$ upper limits.}
\end{center}
\end{table}

Since pulsars' \gr\ emission is stable \citep{2fpsr13}, we checked the
long-term variability of the NE source by calculating its variability 
index TS$_{var}$. Following the procedure introduced in \citet{nol+12},
we derived fluxes of 14 time bins for its \gr\ emission, with 
each bin containing 300-day data. 
If the fluxes are constant, 
TS$_{var}$ would be distributed as $\chi^{2}$ with 13 degrees of freedom. 
Variable sources would be identified with TS$_{var}$ larger than 27.7
(at a 99\% confidence level).
The computed TS$_{var}$ for the NE source is 8.1, indicating that there 
was no significant long-term variability in its \gr\ emission.

\subsubsection{PSR J1715+46}
\label{subsec:second}

\begin{figure}
   \centering
   \includegraphics[width=0.49\textwidth]{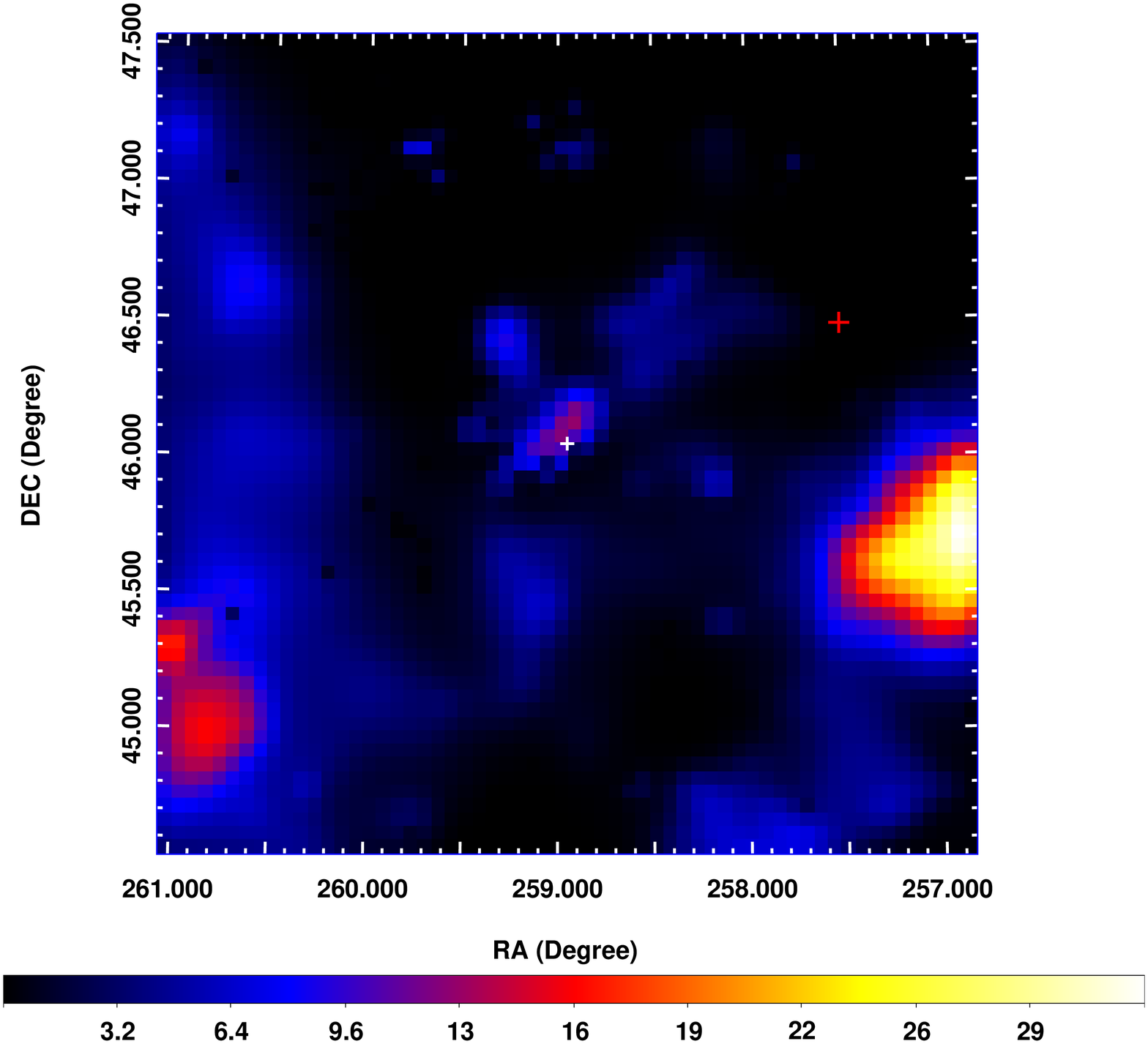}
   \includegraphics[width=0.49\textwidth]{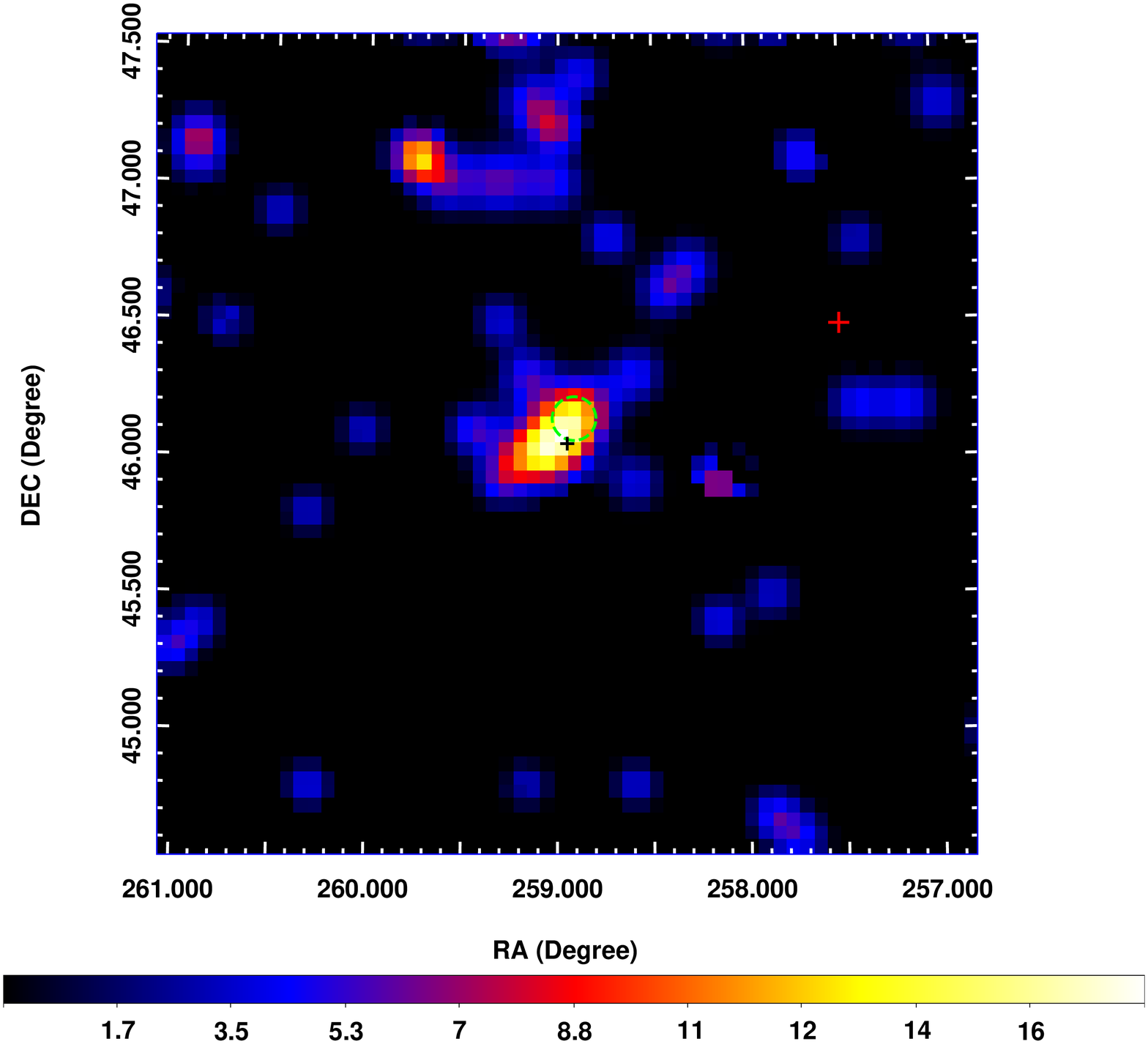}
\caption{TS maps of the $\mathrm{3^{o}\times3^{o}}$ region centered 
at \second\ in the energy ranges of 0.3--500~GeV ({\it left})
and 16--210 GeV ({\it right}). The image scale of the maps is 
0\fdg05 pixel$^{-1}$. There is one catalog source, marked by a red plus, that
was considered and removed.  
The white and black pluses mark the LOTAAS position of \second, with
the length being the positional uncertainty of 3$'$.
The green dashed circle in the right panel is the 2$\sigma$ error circle of 
the best-fit position obtained from \fermi\ analysis, indicating this \gr\
source matches \second\ in position.
}
   \label{fig:j1715}
\end{figure}

The 0.3--500~GeV \gr\ emission near the position of \second\ only 
had TS$\sim$12 
(left panel of Figure~\ref{fig:j1715}), but detailed analysis indicated that
it was more significant in a high energy range of $>$16\,GeV. 
We thus calculated the 16--210 GeV TS map of a $\mathrm{3^{o}\times3^{o}}$ 
region centered at \second\ and a TS value of $\simeq 17$ was found
(right panel of Figure~\ref{fig:j1715}).
We ran \textit{gtfindsrc} to determine the position 
and obtained R.A.=258\fdg91, Decl.=46\fdg14 
(equinox J2000.0) with a 1$\sigma$ nominal uncertainty of 0\fdg05. 
\second\ is 0\fdg09 away from this position. Considering 3$'$ positional
uncertainty of the radio position, the pulsar matches in position with
the \gr\ source.  
Using the \gr\ position, we re-performed the likelihood analysis.
The obtained results were $\Gamma= 1.4\pm$0.3,
$F_{0.3-500}= 5\pm4\times 10^{-11}$ photons~s$^{-1}$\,cm$^{-2}$ in 
the 0.3--500 GeV band (with a TS value of 14), or $\Gamma= 2.2\pm$0.2,
$F_{16-210}= 1.1\pm0.5\times 10^{-11}$ photons~s$^{-1}$\,cm$^{-2}$ 
in the 16--210 GeV band (with a TS value of 16).

Similarly to analysis to the data of \first, we also extracted the \gr\ 
spectrum of \second\ and searched for its long-term variability.
The obtained spectral data points are provided in Table~\ref{tab:spectra}. 
We note that the \gr\ emission was only significantly detected 
in 16.6--38.8 GeV band with a TS value of 17.
The computed TS$_{var}$ for \second\ in 16--210 GeV band is 15.3, 
indicating that there was no significant long-term variability in 
the \gr\ source. 

\section{Discussion}
\label{sec:disc}

Having analyzed \fermi\ LAT data for 70 young pulsars discovered by LOTAAS,
we have found that at the positions of two of them, \first\ and \second, 
there was a \gr\ source respectively. The two sources were faint, 
and the detection significances were only $\simeq 4\sigma$. For the \gr\
source to \second, our spectral analysis showed that the
detection was mostly at the energy range of 16--39 GeV. Such emission
is not consistent with the general \gr\ properties of pulsars, as most of their
spectra have an exponential cutoff at several GeV \citep{2fpsr13}. 
Therefore we only suggest the \gr\ source at the position of \first\ as
a candidate counterpart. In the near future, once the timing solution 
of \first\ is obtained, further data analysis to search for the pulsation signal
at $\gamma$-rays would be able to verify if it is the counterpart.
For \second, we derived its 0.3--500 GeV flux upper limit, which is
$2.6\times 10^{-12}$\,erg\,cm$^2$\,s$^{-1}$.

\begin{figure}
	\centering
	\includegraphics[width=0.50\textwidth]{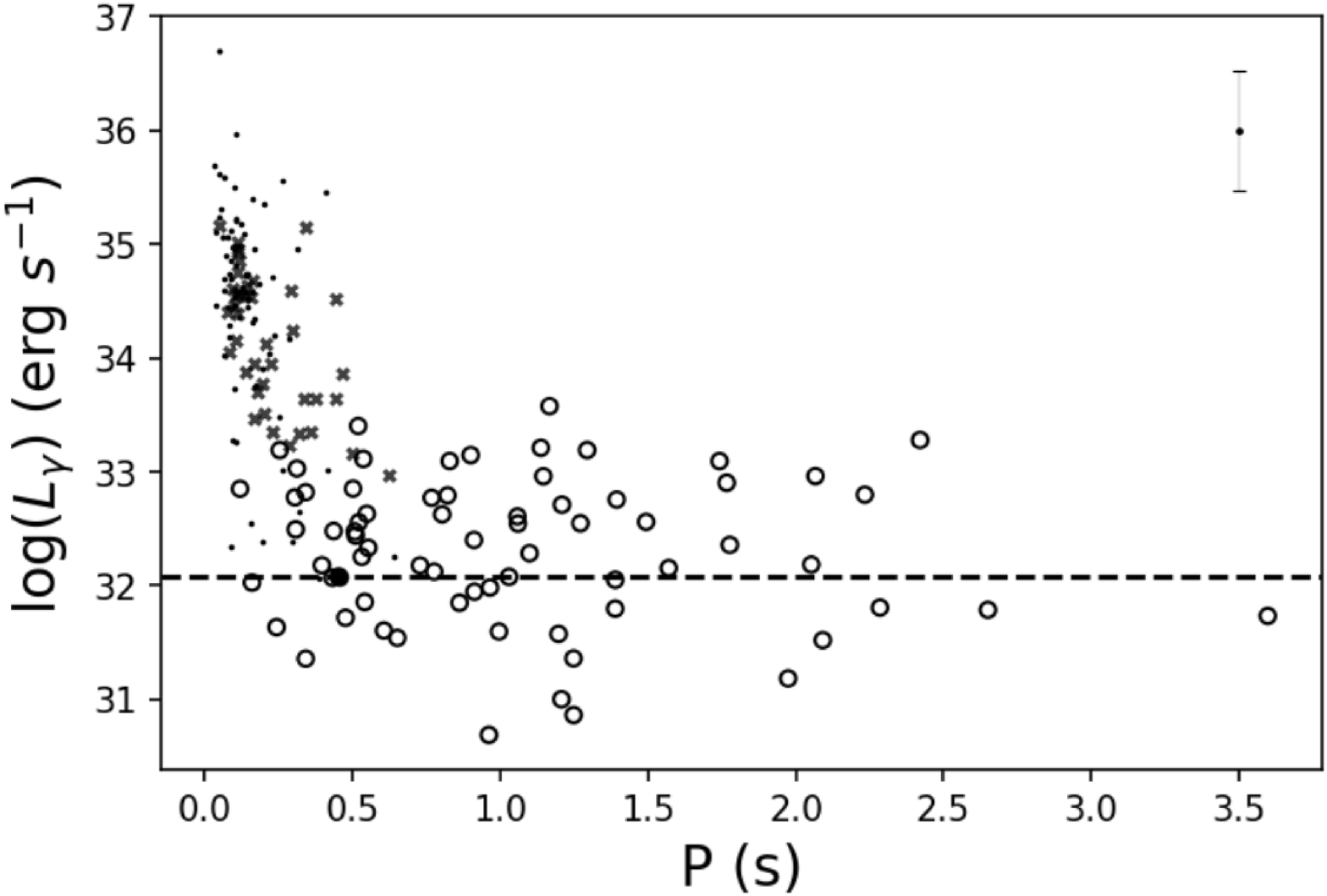}
	\caption{0.3--500 GeV luminosities of the \gr\ pulsars (black dots
 and crosses, where the latter are {\it gu} pulsars with questionable
distances) and luminosity upper limits for the LOTAAS pulsars (open circles). 
\first\ is
marked as a filled circle. The indicative detection limit of the LAT survey
is shown as the dashed line. A 30\% uncertainty is shown in the upper right
conner to help indicate large uncertainties of the luminosities. }
\label{fig:grlum}
\end{figure}

Because of the lack of the period derivatives $\dot{P}$ of the LOTAAS pulsars, 
their
properties such as the spin-down luminosities $\dot{E}$ and surface magnetic 
fields $B$
can not be estimated. In Figure~\ref{fig:grlum}, we show  
$L_{\gamma}$ of the known \gr\ pulsars and the luminosity upper limits
obtained for the LOTAAS pulsars, where for the former see 
appendix~\ref{sec:app}
and for the latter their distances were estimated from the dispersion measures
\citep{san+19}. The former have 
$L_{\gamma} \geq 10^{32}$\,erg\,s$^{-1}$ and spin period $P\leq 0.6$~s, 
where 41 of the {\it gu} pulsars have questionable distance values
(marked with crosses in Figure~\ref{fig:grlum}; see appendix~\ref{sec:app}).
We note that the sensitivity limit of the \fermi\ LAT survey is approximately
$10^{-12}$\,erg\,cm$^{-2}$\,s$^{-1}$ \citep{4fgl20}, and 
assuming 1~kpc distance for a pulsar, the luminosity limit would be
$1.2\times 10^{32}$\,erg\,s$^{-1}$. Thus the \gr\ pulsar detection has
reached this indicative limit (Figure~\ref{fig:grlum}).
For a few LOTAAS pulsars, the upper limits are below $10^{32}$\,erg\,s$^{-1}$, 
and several of them have $P \leq 0.6$~s. The comparison shows that our search
is likely sensitive enough and should have been able to detect some of 
the LOTAAS pulsars if they have \gr\ emission similar to those of 
the \gr\ pulsars.
It is interesting to note that \first\ is right at the range bottom for the
\gr\ pulsars,
which helps not rule out our identification of its \gr\ counterpart.

\begin{figure}
	\centering
	\includegraphics[width=0.50\textwidth]{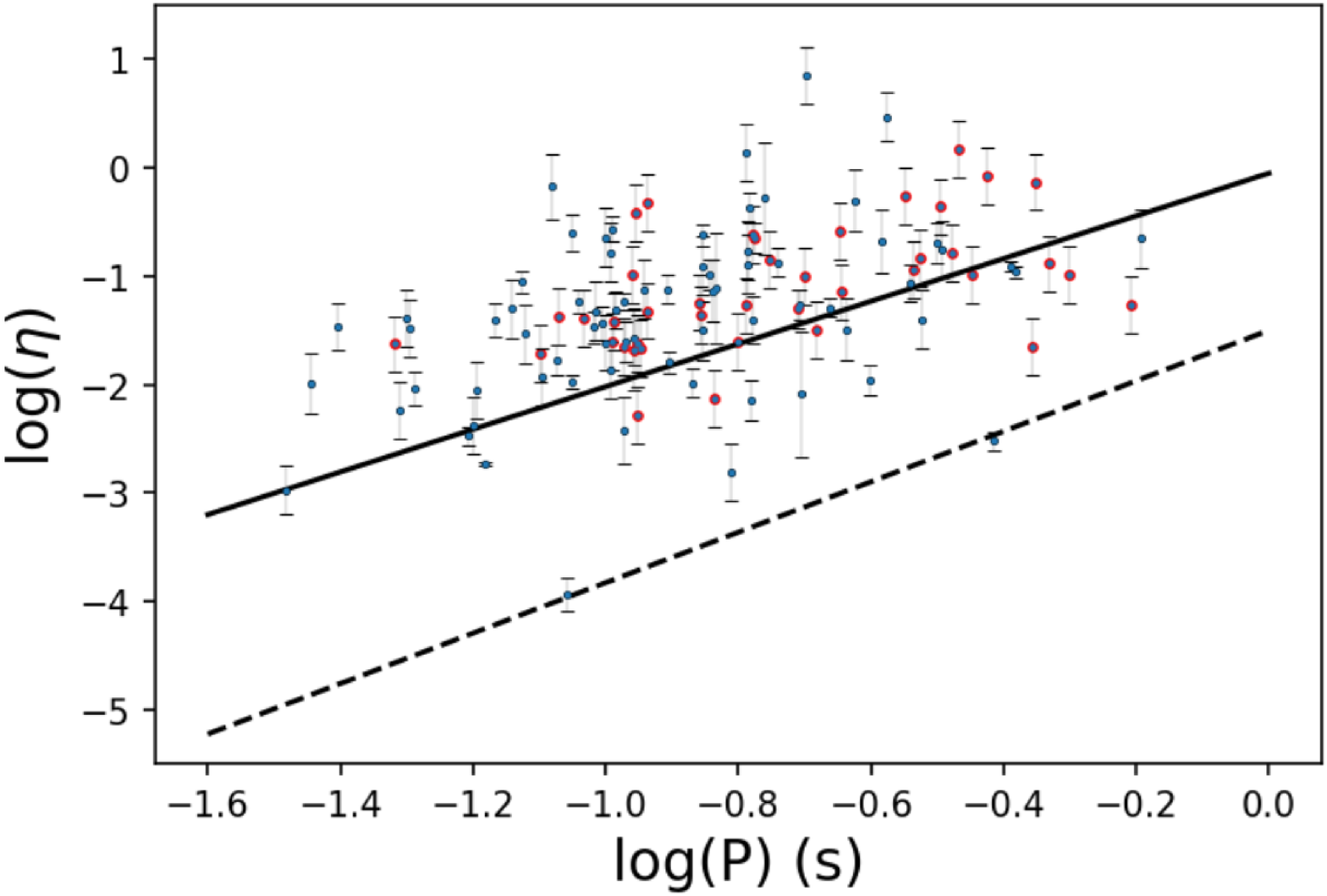}
	\caption{$\log (\eta)$ and $\log (P)$ of 
the \gr\ pulsars  (red data points are {\it gu} pulsars with questionable
distances). A relationship $\log (\eta) = 1.97\log (P) - 0.05$ 
may describe
the data points (solid line). The dashed line is $\log (\eta)=2.33\log (P)-1.5$,
which is a lower limit we define based on these pulsars.}
\label{fig:ep}
\end{figure}

For \gr\ pulsars, relationship $L_{\gamma} \sim \dot{E}^{1/2}$ is often
considered \citep{pie+12,2fpsr13}. Thus \gr\ efficiency 
$\eta = L_{\gamma}/\dot{E} \sim \dot{E}^{-1/2}$. 
Since for pulsars $\dot{E} \sim \dot{P}/P^{3}$, and considering magnetic
dipole radiation from pulsars (e.g., \citealt{lf12}), 
$B\sim (P\dot{P})^{1/2}$, we can find
$\dot{E}\sim B^2/P^{4}$ and $\eta\sim P^2/B$. In Figure~\ref{fig:ep}, 
we show $\log (\eta)$ and $\log (P)$ of the \gr\ pulsars. A possible trend
between them is seen. We test to fit the data points with 
function $\log(\eta)=a\log(P)+b$, and obtain $a\simeq 1.97$ and $b\simeq -0.05$,
 where 41 {\it gu} pulsars with questionable distances were not included
in the fitting.
This possible trend is close to the relationship $\eta\sim P^2$ as well as
$P^{2.33}$ predicted from considering curvature radiation from \gr\ pulsars
(see details in \citealt{kal+19}), while note 
that $B$ is unknown and probably around 10$^{12}$\,G. However, since at any
given $\log(P)$, there is a large scatter in $\log(\eta)$, the fit is not
good (with reduced $\chi^2\sim 30$). We find that we may define a lower limit
line
to the data points with $\log(\eta)=2.33\log(P)-1.5$ (Figure~\ref{fig:ep}), 
for which we specifically require $\eta\sim P^{2.33}$ to match 
the theoretical prediction of curvature radiation from pulsars.

\begin{figure}
	\centering
	\includegraphics[width=0.50\textwidth]{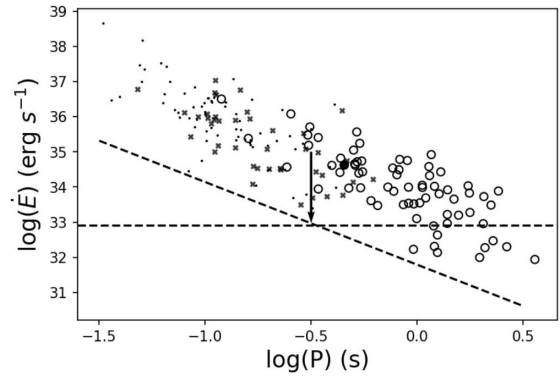}
	\caption{Upper limits on $\dot{E}$ of LOTAAS pulsars (open circles), 
derived from lower limit line $\log(\eta)\geq 2.33log(P) -1.5$ 
(cf. Figure~\ref{fig:ep}). These upper limits could be 100 times smaller 
(indicated by the arrow)
because of the use of the lower limit line. An indicative detection limit
of the LAT survey is shown as the dashed line (see the text for details).
Values of $\dot{E}$ of the \gr\ pulsars (black dots  and crosses) are also 
shown for 
comparison, and a dotted line of $8\times 10^{32}$ erg\,s$^{-1}$ is plotted
to indicate a possible death line for \gr\ emission of pulsars \citep{smi+19}.
The filled circle marks \first.}
\label{fig:edot}
\end{figure}

Using this lower limit, we may set constraints on $\dot{E}$ of the LOTAAS 
pulsars.  Since $\eta= L_{\gamma}/\dot{E}\leq L_{\gamma}^{u}/\dot{E}$, where 
$L_{\gamma}^{u}$ is the \gr\ luminosity upper limits for the LOTAAS pulsars,
we then have $\dot{E}\leq L_{\gamma}^{u}/\eta$. With the lower limit 
relationship given above, the upper limits on $\dot{E}$ of the LOTAAS pulsars
are calculated, which are shown in Figure~\ref{fig:edot}. In this calculation
for \first, the luminosity of the candidate \gr\ counterpart is used. For
comparison, we also include $\dot{E}$ of the \gr\ pulsars in the figure.
The \gr\ pulsars generally have $\dot{E}>10^{33}$\,erg\,s$^{-1}$, and
a death line around this value for \gr\ emission of pulsars has been 
considered \citep{wh11,smi+19}. Most of
the LOTAAS pulsars have upper limits above this $\dot{E}$ value, and are
mixed with the \gr\ pulsars in Figure~\ref{fig:edot}. However because of
the large scatters of $\log(\eta)$, for example, the solid line possibly
describing $\log(\eta)$ and $log(P)$ of the \gr\ pulsars in Figure~\ref{fig:ep}
would increase $\eta$ 100 times and thus lower $\dot{E}$ of the LOTAAS pulsars
100 times correspondingly, which will move most the LOTAAS pulsars below 
$10^{33}$\,erg\,s$^{-1}$. In addition, considering the indicative luminosity
limit of the LAT survey for pulsars and still using this relationship, we have
$\log(\dot{E})=31.79-2.35\log(P)$, which is shown in Figure~\ref{fig:edot}.
Most of the LOTAAS pulsars will also be moved below this detection line.
Therefore it is possible that most of the LOTAAS pulsars could have low 
$\dot{E}$ and thus low \gr\ emission, not to be expected detectable 
with \fermi\ LAT.

As a summary, we have conducted analysis of the \fermi\ LAT data to search
for \gr\ emission from the newly discovered 70 LOTAAS pulsars, and only found
a candidate counterpart to one of them, \first. 
For this pulsar, once its long-term timing solution 
is obtained, a search for its pulsation signal in the \gr\ emission of
the candidate can be conducted for the purpose of verifying if it is the
\gr\ counterpart.
We provide the 0.3--500~GeV
flux upper limits for the non-detections. By comparing the LOTAAS pulsars with
the known \gr\ pulsars, we estimate the $\dot{E}$ upper limits for the LOTAAS
pulsars. The upper limits are not very constraining, and it is likely that
most of the LOTAAS pulsars have low, $<10^{33}$\,erg\,s$^{-1}$ $\dot{E}$ and 
are not expected to have detectable \gr\ emission.

\begin{acknowledgements}
We thank anonymous referees for very helpful suggestions.
This research made use of the High Performance Computing Resource in the Core
Facility for Advanced Research Computing at Shanghai Astronomical Observatory.
This research was supported by the National Program on Key Research 
and Development Project (Grant No. 2016YFA0400804) and
the National Natural Science Foundation of China (11633007, U1738131). Z.W.
acknowledges the support by the Original Innovation Program of the
Chinese Academy of Sciences (E085021002).
\end{acknowledgements}

\appendix

\section{Known $\gamma$-ray pulsars}
\label{sec:app}

There are 121 young \gr\ pulsars listed in 4FGL \citep{4fgl20}. 
Searching in the Australia Telescope National Facility (ATNF) pulsar 
catalog \citep{man+05} and references provided at MCG, we found distance values for 112 of them, which are given in Table~\ref{tab:gr}. 
 However it can be noted that 53 of 121 \gr\ pulsars are marked with 
`{\it gu}' at MCG,  which indicates that these pulsars were discovered in 
LAT data and/or using an LAT seed position. As most of the {\it gu} pulsars 
have not been detected at radio frequencies, they are probably assigned with 
`heuristic' distances (see \citealt{2fpsr13}). We searched references for
{\it gu} pulsars in Table~\ref{tab:gr} and found no radio detection reported 
for 41 of them.
To be cautious, we marked these 41 {\it gu} pulsars in Table~\ref{tab:gr} with 
$g$, indicating that their distance values are questionable.

Using the spectral parameters given in 4FGL, we derived 0.3--500 GeV flux 
$F_{0.3-500}$.  For the pulsars without
distance uncertainties, we assumed 30\% uncertainties (e.g., \citealt{cam+09}).
We calculated the luminosities from $L_{\gamma}=4\pi d^2f_{\Omega}F_{0.3-500}$,
where $d$ is the distance and $f_{\Omega}$ is the beam correction factor.
Following \citet{2fpsr13}, we adopted $f_{\Omega}=1$.
\begin{ThreePartTable}
\begin{TableNotes}
\item[$a$] Distances without errors are assumed to have 30\% uncertainties.
\item[$g$] \gr\ pulsars without (reported) radio detection.
\end{TableNotes}
\begin{longtable}{lcccccc}
\caption{0.3--500 GeV fluxes and luminosities for 112 $\gamma$-ray pulsars.}
\label{tab:gr}\\
\hline
Source & P & Distance$^a$ &$ \dot{E}$/10$^{33}$& Flux/10$^{-12}$ & Luminosity \\
name &(s)&(kpc)&(erg\,s$^{-1}$)& (erg\,cm$^{-2}$\,s$^{-1}$) & (10$^{33}$erg\,s$^{-1}$) \\ 
\hline
\endfirsthead
\hline
Source & P & Distance$^{a}$ &$ \dot{E}$/10$^{33}$& Flux/10$^{-12}$ & Luminosity \\     
name & (s) & (kpc) & (erg\,s$^{-1}$) & (erg\,cm$^{-2}$\,s$^{-1}$) & (10$^{33}$erg\,s$^{-1}$) \\ 
\hline
\endhead
\hline
\endfoot
\hline
\insertTableNotes
\endlastfoot
J0002$+$6216$^g$ & 0.12 & 6.36 & 150&  14.7$\pm$1.2& 71$\pm$43 \\

J0007$+$7303& 0.32 &1.4$\pm$0.3 &450& 387.2$\pm$3.1 & 91$\pm$39 \\

J0106$+$4855& 0.083 &3.1$\pm$1.1&29 & 17.5$\pm$0.8 & 20$\pm$14\\

%%J0117$+$5914& 0.10 &1.77&220 & 0.4$\pm$0.7 & 0.14$\pm$0.28\\

J0205$+$6449& 0.066 &1.95$\pm$0.04&27000 & 40.8$\pm$1.1&49.9$\pm$2.5\\

J0248$+$6021& 0.22 &2.0$\pm$0.2&210 & 22.5$\pm$1.3&10.8$\pm$2.2\\

J0357$+$3205$^g$& 0.44 &0.83&5.9 & 53.1$\pm$1.0 & 4.4$\pm$2.6\\

J0359$+$5414$^g$& 0.079 &3.45&1300 & 17.7$\pm$1.3 & 25$\pm$15\\

J0514$-$4408& 0.32 &0.97&2.5 & 3.9$\pm$0.3 & 0.43$\pm$0.27\\

J0534$+$2200& 0.033 &2$\pm$0.5&460000 & 1000$\pm$17& 490$\pm$240\\

J0540$-$6919&0.051 &49.7&150000 & 17.0$\pm$1.1&5030$\pm$3034\\

J0554$+$3107$^g$& 0.46 &1.9&56 & 17.0$\pm$0.7 & 7.3$\pm$4.4\\

J0622$+$3749$^g$& 0.33 &1.6&27 & 14.3$\pm$0.6 & 4.4$\pm$2.6\\

J0631$+$0646$^g$& 0.11 &4.58& 100 & 15.3$\pm$1.4&38$\pm$23\\

J0631$+$1036& 0.29 &1.0$\pm$0.2&170 & 27.9$\pm$1.2 & 14.7$\pm$5.9\\

J0633$+$0632$^g$& 0.30 &1.35& 120 & 80.9$\pm$2.3 & 18$\pm$11\\

J0633$+$1746& 0.24 &0.25$\pm$0.08& 32 &  3700$\pm$11 & 16$\pm$10\\

J0659$+$1414& 0.38 &0.29$\pm$0.03& 38 & 11.3$\pm$0.4 & 0.114$\pm$0.024\\

J0729$-$1448& 0.25 &3.5$\pm$0.4& 280 &  3.6$\pm$0.8 & 3.1$\pm$1.0\\

%%J0734$-$1559& 0.16 & &130 &4384&31.4$\pm$1.5& \\

J0742$-$2822& 0.17 &2.1$\pm$0.5& 140& 11.6$\pm$1.0 & 5.5$\pm$2.7\\

J0835$-$4510& 0.089 &0.28$\pm$0.02& 6900 & 7700$\pm$27 & 72$\pm$10 \\

J0908$-$4913& 0.11 &2.6$\pm$0.9&490 & 15.5$\pm$2.5 & 2.5$\pm$1.3\\

%%J0922$+$0638& 0.43 &1.1&6.8 &  1.0$\pm$0.2 &0.15$\pm$0.10\\

J0940$-$5428& 0.088 &3.0$\pm$0.5& 1900 & 12.6$\pm$.9 & 0.218$\pm$0.074\\

J1016$-$5857& 0.11 &2.9$\pm$0.6& 2600 & 54.2$\pm$3.6 & 65$\pm$27\\

J1019$-$5749& 0.16 &10.91& 180 & 17.3$\pm$2.3 & 250$\pm$150\\

J1023$-$5746$^g$& 0.11 &2.08& 11000 & 111.7$\pm$6.9 & 58$\pm$35\\

J1028$-$5819& 0.091 &2.3$\pm$0.3& 840 & 204.2$\pm$4.7 & 49$\pm$13\\

J1044$-$5737$^g$& 0.14 &1.9& 800 & 82.3$\pm$2.6 & 36$\pm$21\\

J1048$-$5832& 0.12 &2.7$\pm$0.4& 2000 & 150.4$\pm$3.3 & 150$\pm$91\\

J1055$-$6028& 0.10 &3.83& 1200 & 16.6$\pm$1.7 & 29$\pm$18\\

J1057$-$5226& 0.20 &0.3$\pm$0.2& 30 & 252.8$\pm$2.3 & 0.24$\pm$0.33\\

J1057$-$5851$^g$& 0.62 &0.8& 17 & 12.01$\pm$0.88 & 0.92$\pm$0.56\\

J1105$-$6037$^g$& 0.19 &1.53&120 & 21.4$\pm$1.7 & 6.0$\pm$3.6\\

J1105$-$6107& 0.063 &5$\pm$1&2500 & 15.7$\pm$2.0 & 10.4$\pm$4.4\\

%%J1111$-$6039& 0.11 & &  &3217&49.1$\pm$1.9& \\

J1112$-$6103& 0.064 & 4.5 & 4500 & 16.4$\pm$2.8 & 40$\pm$25\\

J1119$-$6127& 0.41 &8.4$\pm$0.4&2300 & 33.5$\pm$2.1 & 280$\pm$32\\

J1124$-$5916& 0.14 &4.8$\pm$0.7&12000 & 41.6$\pm$1.6 & 120$\pm$37\\

%%J1135$-$6055& 0.11 & & 2100 &2878&33$\pm$1.1& \\

J1151$-$6108& 0.10 &2.22& 390 & 9.04$\pm$0.93 & 5.3$\pm$3.2\\

J1208$-$6238$^g$& 0.44 &3& 1500 & 30.8$\pm$2.5 & 33$\pm$20\\

J1253$-$5820& 0.26 &1.64& 5 & 3.26$\pm$0.99 & 1.05$\pm$0.71\\

J1341$-$6220& 0.19 &12.6& 1400 & 16.9$\pm$2.8 & 320$\pm$199\\

J1350$-$6225$^g$& 0.14 &1.3& 130 & 36.9$\pm$2.3 & 7.4$\pm$4.5\\

J1357$-$6429& 0.17 &2.5$\pm$0.5& 3100 & 19.3$\pm$1.3 & 22.22$\pm$0.90\\

J1410$-$6132& 0.050 &15.6$\pm$4.2& 10000 & 18.9$\pm$4.7 & 410$\pm$240\\

J1413$-$6205$^g$& 0.11 &2.15& 830 & 155.3$\pm$4.8 & 86$\pm$52\\

J1418$-$6058$^g$& 0.11 &1.6$\pm$0.7& 4900 & 240$\pm$12 & 100$\pm$91\\

J1420$-$6048& 0.068 &5.6$\pm$0.9& 10000 & 100$\pm$11 & 390$\pm$130\\

J1422$-$6138$^g$& 0.34 &4.8& 96 & 51.9$\pm$2.7 & 140$\pm$86\\

J1429$-$5911$^g$& 0.12 &1.95& 770& 81.0$\pm$4.3 & 37$\pm$22 \\

J1459$-$6053$^g$& 0.10 &1.84& 910 & 86.2$\pm$2.3 & 35$\pm$21\\

J1509$-$5850& 0.089 &2.6$\pm$0.5& 520& 97.2$\pm$2.8 & 130$\pm$51\\

J1522$-$5735$^g$& 0.10 &2.1& 1200 & 55.8$\pm$2.6 & 29$\pm$18\\

J1528$-$5838$^g$& 0.36 &1.1& 22 & 15.4$\pm$1.3 & 2.2$\pm$1.4\\

J1531$-$5610& 0.084 &2.1$\pm$0.3&900 & 15.7$\pm$1.9 & 15.1$\pm$9.3\\

%%J1615$-$5137& 0.18 & &73 &954&33.86$\pm$0.27& \\

J1614$-$5048& 0.23 &5.15 & 1600 & 16.2$\pm$4.4 & 52$\pm$34\\

J1623$-$5005$^g$& 0.085 &1.3&270 & 5.6$\pm$3.3 & 11.4$\pm$6.9\\

J1624$-$4041$^g$& 0.17 &1.8& 39 & 23.1$\pm$1.1 & 8.9$\pm$5.4\\

%%J1641$-$5317& 0.18 & &27&856&13.1$\pm$0.8&  \\

J1648$-$4611& 0.16 &4.5$\pm$0.7& 210& 37.7$\pm$2.2 & 90$\pm$28 \\

J1702$-$4128& 0.18 &4.8$\pm$0.6& 340& 24.2$\pm$4.0 & 45$\pm$14 \\

J1705$-$1906& 0.30 &0.75& 6.1 & 3.64$\pm$0.46 & 0.24$\pm$0.15\\

J1709$-$4429& 0.10 &2.3$\pm$0.3& 1153.2$\pm$8.0 & 930$\pm$240\\

%%J1714$-$3830& 0.084 & & &3132&73.13$\pm$0.058&  \\

J1718$-$3825& 0.075 &3.6$\pm$0.4&1300& 79.7$\pm$4.8 & 120$\pm$27 \\

J1730$-$3350& 0.14 &3.49 & 1200 & 26.1$\pm$4.9 & 38$\pm$24 \\

J1732$-$3131& 0.20 &0.6$\pm$0.1&150 & 166.7$\pm$3.4 & 8.2$\pm$2.7\\

J1739$-$3023& 0.11 &3.07& 300 & 20.0$\pm$3.2 & 23$\pm$14\\

J1740$+$1000& 0.15 &1.23& 230 & 1.95$\pm$0.37 & 0.35$\pm$0.22\\

J1741$-$2054& 0.41 &0.38$\pm$0.02& 9.5 &  96.3$\pm$1.9 & 1.04$\pm$0.11\\

J1746$-$3239$^g$& 0.20 &0.8& 33 & 42.5$\pm$2.6 & 3.3$\pm$2.0   \\

J1747$-$2958& 0.10 &4.8$\pm$0.8& 2500 & 121.3$\pm$6.7 & 92.1$\pm$31.1\\

%%J1757$-$2421& 0.23 &3.12& 40 & 11.82$\pm$0.04 &13.8$\pm$8.3\\

J1801$-$2451& 0.12 &5.2$\pm$0.5& 2600 & 24.2$\pm$2.6 & 41.8$\pm$9.2\\

J1803$-$2149$^g$& 0.11 &1.3& 640 & 69.8$\pm$3.8 & 14.1$\pm$8.5\\

J1809$-$2332& 0.15 &1.7$\pm$1.0& 430 & 362.3$\pm$6.2 & 34$\pm$39\\

J1813$-$1246$^g$& 0.048 &2.63& 6200 & 177.2$\pm$4.1 & 150$\pm$88\\

%%J1817$-$1742& 0.15 & & 240 &348&21.18$\pm$0.050& \\

J1826$-$1256& 0.11 &1.55& 3600 & 340.0$\pm$8.8 & 98$\pm$59\\

J1827$-$1446$^g$& 0.50 &0.7&14 & 24.7$\pm$1.7 & 1.44$\pm$0.88\\

J1828$-$1101& 0.072 & 4.77 & 1600 & 29.6$\pm$5.0 & 80.3$\pm$50.1\\

%%J1831$-$0952& 0.067 &3.68& 1100& 8.2$\pm$0.1&13.3$\pm$8.0\\

J1833$-$1034& 0.062 &4.7$\pm$0.4& 34000& 56.4$\pm$4.6 & 110$\pm$21 \\

J1836$+$5925& 0.17 &0.5$\pm$0.3& 11 & 537.7$\pm$2.9 & 5.8$\pm$7.0\\

J1837$-$0604& 0.096 &4.77& 2000 & 35.2$\pm$5.8 & 96$\pm$60\\

J1838$-$0537$^g$& 0.15 &2& 5900 & 9.1$\pm$1.0 & 44$\pm$27\\

J1844$-$0346$^g$& 0.11 &4.3& 4300 & 41.8$\pm$2.8 & 92$\pm$56\\

J1846$+$0919$^g$& 0.23 &1.53& 34 & 32.0$\pm$1.0 &8.9$\pm$5.4\\

J1853$-$0004& 0.10 &5.34& 210 & 10.2$\pm$1.7 & 35$\pm$22\\

J1857$+$0143& 0.14 &4.57& 450 & 22.3$\pm$3.6 & 56$\pm$35\\

J1906$+$0722$^g$& 0.11 &1.91& 1000 & 55.5$\pm$5.6 & 24$\pm$15\\

J1907$+$0602& 0.11 &3.2$\pm$0.3& 2800 & 240.9$\pm$5.2 & 160$\pm$31\\

J1913$+$0904& 0.16 &3& 160 & 19.3$\pm$1.9 & 21$\pm$13\\

J1913$+$1011& 0.036 & 4.61 & 2900 & 11.5$\pm$2.5 & 29$\pm$19\\

J1925$+$1720& 0.076 &5.06& 950 & 9.1$\pm$1.5 & 28$\pm$17\\

J1932$+$1916$^g$& 0.21 &1.5& 410 & 48.9$\pm$3.5 & 13.1$\pm$7.9\\

J1932$+$2220& 0.14 &10.9& 750 & 3.84$\pm$0.98 & 55$\pm$36\\

J1935$+$2025& 0.08 &4.6& 4600 & 21.5$\pm$1.9 & 54$\pm$33\\

J1952$+$3252& 0.039 &2.0$\pm$0.5& 3700& 118.4$\pm$2.3 & 120$\pm$64 \\

J1954$+$2836$^g$& 0.092 &1.96& 1000 & 8.8$\pm$2.9 & 41$\pm$24\\

J1957$+$5033$^g$& 0.37 &1.36& 5.3 & 20.1$\pm$0.63 & 4.4$\pm$2.7\\

J1958$+$2846$^g$& 0.29 &1.95& 340& 86.3$\pm$2.8 & 39$\pm$24 \\

J2006$+$3102& 0.16 &6.03& 220 & 8.7$\pm$1.4 & 38$\pm$24\\

J2017$+$3625$^g$& 0.17 &0.656& 12 & 56.8$\pm$3.8 & 2.9$\pm$1.8\\

J2021$+$3651& 0.10 &10$\pm$2& 3400& 424.1$\pm$5.4 & 160$\pm$66 \\

J2022$+$3842& 0.08 &4.6& 4700 & 14.5$\pm$2.1 & 170$\pm$110\\

J2021$+$4026& 0.27 &1.5$\pm$0.4& 120 & 645.5$\pm$8.8 & 360$\pm$190\\

J2028$+$3332$^g$& 0.18 &0.9& 35 & 51.5$\pm$1.5 & 5.0$\pm$3.0\\

J2030$+$3641& 0.20 &3$\pm$1& 32 & 39.2$\pm$1.8 & 230$\pm$140\\

J2030$+$4415$^g$& 0.23 &0.72& 32 & 36.7$\pm$1.5 & 2.3$\pm$1.4 \\

J2032$+$4127& 0.14 &3.7$\pm$0.6&270 & 130.7$\pm$3.8 & 27.6$\pm$9.0\\

J2043$+$2740& 0.096 &1.8$\pm$0.3&56 & 7.22$\pm$0.52 & 1.89$\pm$0.64\\

J2055$+$2539$^g$& 0.32 &0.62& 4.9 & 46.59$\pm$0.96 & 2.1$\pm$1.3\\

J2111$+$4606$^g$& 0.16 &2.7& 1400& 39.6$\pm$1.2 & 35$\pm$21 \\

J2139$+$4716$^g$& 0.28 &0.8& 3.1 &  22.48$\pm$0.98 & 1.7$\pm$1.0\\

J2208$+$4056& 0.64 &0.75& 0.81 &  2.68$\pm$0.43 & 0.18$\pm$0.11\\

J2229$+$6114& 0.052 &0.8$\pm$0.15& 22000& 187.8$\pm$2.7 & 200$\pm$76 \\

J2238$+$5903$^g$& 0.16 &2.83&890 &  49.7$\pm$1.6 & 47.6$\pm$28.6\\

J2240$+$5832& 0.14 &7.7$\pm$0.7& 220 & 8.4$\pm$1.0 & 53$\pm$12\\
\end{longtable}

\end{ThreePartTable}

%%%\bibliography{lofar}

\end{document}